\documentclass[twocolumn,longbibliography,aps,prx,floatfix,
preprintnumbers]{revtex4-2}

\usepackage[urlcolor=blue,colorlinks=true,citecolor=blue,linkcolor=blue,
bookmarks=false]{hyperref}

\usepackage{graphicx}
\usepackage{bm}
\usepackage{xcolor}
\usepackage{amsmath}
\usepackage{amssymb}

\usepackage{lipsum}
\usepackage{gensymb}
\usepackage{orcidlink}
\usepackage{multirow}

\begin{document}
	\title{Anomalous magnetotransport in single-crystalline half-Heusler antiferromagnet ErPdSb}
	\author{Abhinav Agarwal\orcidlink{0009-0006-6332-9374}}
	\author{Shovan Dan \orcidlink{0000-0001-7763-2573}}
    \author{Maciej J. Winiarski\orcidlink{0000-0001-6935-4825}}
	\author{Orest Pavlosiuk\orcidlink{0000-0001-5210-2664}}
    \author{Piotr Wiśniewski\orcidlink{0000-0002-6741-2793}} 
	\author{Dariusz Kaczorowski\orcidlink{0000-0002-8513-7422}} 
    \email{d.kaczorowski@intibs.pl}
    
	\affiliation{Institute of Low Temperature and Structure Research, Polish Academy of Sciences, Okólna 2, 50-422 Wrocław, Poland}
	
	\begin{abstract}
    We report the thermodynamic and magnetotransport properties of a half-Heusler antimonide ErPdSb, studied on single-crystalline samples in wide ranges of temperature and magnetic fields. The compound was found to order antiferromagnetically at 1.2 K. In the paramagnetic state, it shows semimetallic behavior with a broad hump in the temperature-dependent electrical resistivity around 70~K. The results of \textit{ab~initio} calculations of the electronic structure of ErPdSb indicated a bulk insulating nature. In small magnetic fields, the magnetoresistance is driven by weak antilocalization effect, while in strong fields, it is negative and describable by de Gennes - Friedel formalism. The Hall effect data indicated that holes are the dominant charge carriers. At 2 K, the Hall conductivity exhibits a sizable anomalous contribution, which is obscured by multiband effects at higher temperatures. The angular magnetoresistance shows unusual features as functions of magnetic field and temperature, pointing to a possible field-induced reconstruction of the Fermi surface.
	\end{abstract}
	\maketitle

\section{Introduction}
In the last two decades, extensive research has been done on rare-earth (RE) based half-Heusler (HH) phases that provide a plethora of interesting magnetic \cite{Chen2021}, heavy-fermion \cite{Canfield1991, Movshovich1994}, superconducting \cite{Meinert2016, Tafti2013}, topological \cite{Lin2010, Singh2020}, and/or thermoelectric \cite{Mastronardi1999} properties. Moreover, several representatives of the HH family have been recognized as promising candidates for practical spintronic applications \cite{Chris2016}. 
Recently, a group of RETBi (T = Pd, Pt) bismuthides has attracted particular attention because of their unique thermodynamic and transport behaviors, linked to their nontrivial electronic structures, which exhibit a band inversion effect. The majority of these compounds are ordered antiferromagnetically (AFM) at low temperatures \cite{Nakajima2015, Pavlosiuk2019, Pavlosiuk2016, Gofryk2011, Pan2013, Zhu2023, Hirschberger2016, Shekhar2018, Pavlosiuk2020, Zhang2020}, and some of them exhibit superconductivity with inherent parity mixing \cite{Goll2008,Pavlosiuk2021,Xu2014,Ishihara2021}. In a few materials, viz. REPdBi (RE = Sm, Dy, Ho, Er) \cite{Nakajima2015, Pavlosiuk2016, Pan2013}, both cooperative phenomena coexist, making them excellent candidates for comprehensive studies on the interplay between magnetism, superconductivity, and nontrivial topology. A large anomalous Hall effect, negative magnetoresistance, and the planar Hall effect are widely reported in this class of materials. It was shown that these anomalous physical properties are linked with field-induced Weyl nodes \cite{Hirschberger2016, Bhattacharya2023}, spin-chirality \cite{Suzuki2016}, chiral magnetic anomaly (CMA) \cite{Kumar2018,Pavlosiuk2019}, and other features characteristic of non-trivial electronic band structures \cite{Zhu2023, Lu2025}.

In contrast to the rather widely investigated RETBi series, their antimony-containing counterparts have attracted much less attention so far. This is principally because of the expectation of a lower spin-orbit coupling (SOC) strength after the replacement of Bi atoms with atomic number Z = 83 by Sb atoms with Z = 51 (SOC$\propto$ $\mathrm{Z^4}$). As a result, the SOC interaction may become too weak to induce inversion of the electronic bands in the RETSb compounds from a topologically trivial order $\Gamma_6-\Gamma_8-\Gamma_7$ into a non-trivial one $\Gamma_8-\Gamma_6-\Gamma_7$ \cite{Chadov2010, Yan2014}.

Almost all magnetic and transport data of RETSb available in the literature have been obtained on polycrystalline samples \cite{Malik1991, Gofryk2005, Gofryk2007, Shekhar2012, Mukhopadhyay2018}, with the rare exception of TmPdSb studied recently on high-quality single crystals \cite{Dan2024}. Interestingly, the latter compound turned out to be an insulator with an indirect gap, highly unusual electronic bands inversion, and metallic surface states, which was a completely new finding for the RE-based HH compounds \cite{Dan2024}.

Polycrystalline ErPdSb was reported to be a Curie-Weiss paramagnet with no long-range magnetic order down to 1.9 K \cite{Kaczorowski2005, Gofryk2005, Gofryk2007, Mukhopadhyay2018}. The compound was found to be a small-gap semiconductor with electrical transport governed by multiple-band contributions, and promising thermoelectric and magnetocaloric characteristics at elevated temperatures \cite{Gofryk2007, Mukhopadhyay2018}.

In this work, we reinvestigated ErPdSb on single-crystalline specimens, focusing at the magnetic behavior at temperatures below 2 K, and the electrical magnetotransport (angle-dependent magnetoresistance, Hall effect) in wide ranges of temperature and magnetic field strength.

\section{Methods}
\subsection{Experimental details}
Single crystals of ErPdSb were grown from Bi flux. High-purity elements Er, Pd, Sb and Bi were taken in ratio 1:1:1:30, placed in an alumina crucible, and sealed in an evacuated quartz tube. The ampule was slowly heated to 1050 $^{^\circ}$C, held at this temperature for 30 hours, and then slowly cooled to 650 $^{^\circ}$C at a rate of 2 $^{^\circ}$C per hour. Subsequently, the tube was removed from the furnace and centrifuged to remove the flux. As the product, several silver shiny cube-shaped crystals were isolated, with a typical size of about $1 \times 1 \times 1$ mm$^3$.

The chemical composition of the obtained crystals was checked by means of energy-dispersive x-ray spectroscopy (EDS) performed  using a FEI scanning electron microscope equipped with a Genesis XM4 EDS probe. The EDS results indicated the expected equiatomic stoichiometry and homogeneous composition of the samples. The crystal structure was 

In order to verify the crystal symmetry of the crystals, a small fragment was crumbled from a larger piece and examined on an Oxford Diffraction X'calibur four-circle single-crystal X-ray diffractometer equipped with a CCD Atlas detector. The experiment yielded a cubic unit cell with the parameter $a$ = 6.472(3) Å, in agreement with the literature data \cite{Kaczorowski2005, Gofryk2005, Gofryk2007, Mukhopadhyay2018}. The crystal structure refinement (see the Supplementary Material, SM \cite{SuppMat}) confirmed the non-centrosymmetric structure of the MgAgAs type (space group $F\bar43m$, No. 216), with the constituent atoms occupying the following Wyckoff positions: Er at 4$a$ (0,0,0), Sb at 4$b$ (0.5,0.5,0.5), and Pd at 4$c$ (0.25,0.25,0.25). The crystals selected for physical properties measurements were oriented employing backscattering Laue X-ray diffraction technique implemented in a Proto Manufacturing Laue-COS camera. Then, the specimens were cut with respect to their (001) planes using a wire saw.

Magnetic measurements were carried out in the temperature range $T$ = 0.5--300 K and in magnetic fields $B$ up to 7 T using a Quantum Design MPMS-XL superconducting quantum interference device magnetometer equipped with an iQuantum He3 refrigerator. The heat capacity was measured from 300 K down to 0.4 K in fields $B$ = 0, 0.5, and 1 T using a relaxation method implemented in a Quantum Design PPMS-9 platform with a He-3 insert. The specimen was mounted on the experimental puck using Apiezon N grease. Electrical transport studies were performed in the interval $T$ = 2--300 K in magnetic fields up to 9 T employing an ac four-point technique and the same PPMS-9 equipment. The measurements were conducted on a bar-shaped sample with a cross-sectional area of 0.61~mm$^2$ and a voltage lead separation of 1.1~mm. Electrical contacts were made using silver epoxy paste and silver wires with a diameter of 50~$\mu$m. The raw magnetoresistance and Hall resistivity data were symmetrized and antisymmetrized, respectively, using the standard protocol.

\subsection{Computational details}
Electronic structure calculations were performed using the VASP package \cite{VASP1,VASP2,VASP3}. The generalized gradient approximation (GGA \cite{GGA}) and meta-GGA modified Becke-Johnson approach (MBJGGA \cite{MBJ} were employed in the fully relativistic mode (with included spin-orbit coupling, SOC). The Er $4f$ states were treated as core states. The plane-wave cut-off was set to 600 eV. The 16$\times$16$\times$16 {\bf k}-point mesh was employed. The Brillouin zone of the face-centered cubic (fcc) primitive cell of ErPdSb is shown in the inset of Fig. \ref{fig:BAND}(a).

\section{Results}
\subsection{Electronic structure}
\begin{figure}[t]
    \centering
    \includegraphics[width=0.8\linewidth]{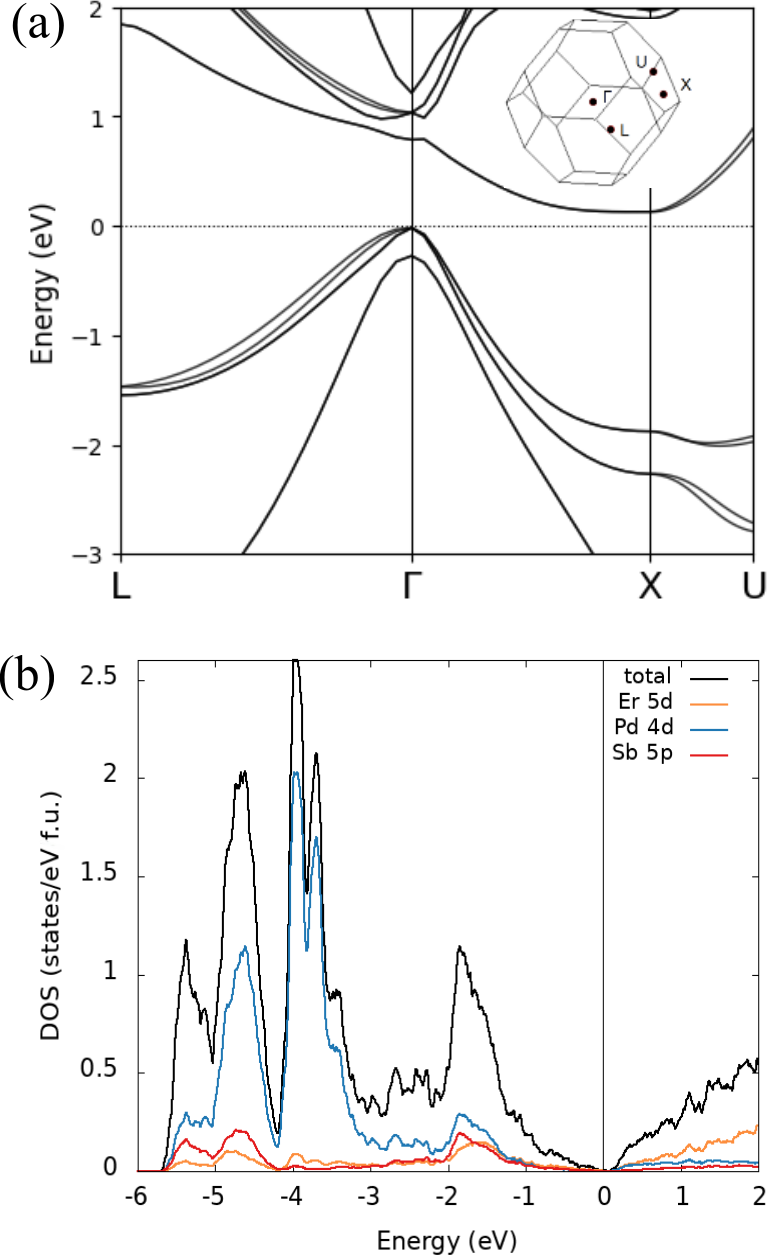}
    \caption{(a) Electronic band structure calculated using the MBJGGA approach. The inset shows a Brillouin zone of the fcc primitive cell of ErPdSb. (b) Total and partial density of states.}
    \label{fig:BAND}
\end{figure}

Fig. \ref{fig:BAND} (a) displays the band structure of ErPdSb calculated using the MBJGGA technique and the experimental lattice parameter. In both GGA and MBJGGA approaches, an indirect band gap of about 0.15 eV was found for the $\Gamma$-$X$ direction. The shape of the valence band maximum (VBM) suggests rather small effective masses of hole carriers, whereas the conduction band minimum is relatively flat in some directions in the Brillouin zone. The splitting between the heavy- and light-hole bands found for ErPdSb equals 0.3 eV, that is only slightly larger than that reported for YPdSb (0.26~eV~\cite{Winiarski2019}). Strong SOC in RE-based HH phases leads to significant splittings of the valence and conduction bands along the $L$-$\Gamma$ and $X$-$U$ lines. Similar overall shapes of bands were recently published for TmPdSb \cite{Dan2024} and are common in Pd- and Ni-bearing HH antimonides with enhanced thermoelectric performance \cite{Winiarski2018, Winiarski2019}.

Interestingly, the GGA calculations (not shown) indicated ErPdSb to be a topologically trivial material without band inversion, where VBM is dominated by the Sb $5p$ states and a conduction band maximum (CBM) of clear $s$-type character. However, the MBJGGA calculations revealed a conduction band inversion, similar to that reported recently for TmPdSb \cite{Dan2024}. In ErPdSb, the $s$-type $\Gamma_6$ states were found to reside 1 eV above the Fermi level, i.e. above the valence bands dominated by the Sb $5p$ states. It is not clear, how such an unusual order of bands may influence the transport properties of a material.

It is also worth considering that the MBJ potential was developed for simple semiconductors and oxides \cite{MBJ}. The electronic structure of transition metal - and RE - based HH materials is often complex and untypical. For instance, the full potential calculations of the electronic structure in YPdSb yielded even a narrower MBJ band gap than the GGA one \cite{Winiarski2019}, which is fairly puzzling because the MBJ approach usually results in band gap widening \cite{TI_1, TI_2}.
  
The total density of states (DOS) in the valence region of ErPdSb, depicted in Fig. \ref{fig:BAND}(b), is dominated by the Pd $4d$ states with small contributions from the Er $5d$ and Sb $5p$ states. These contributions become crucial at a characteristic peak at about 1.5 eV below $E_{\rm F}$. The conduction band region is dominated by the unoccupied Er $5d$ states.

\subsection{Thermodynamic properties}
\begin{figure}
    \centering
    \includegraphics[width=1\linewidth]{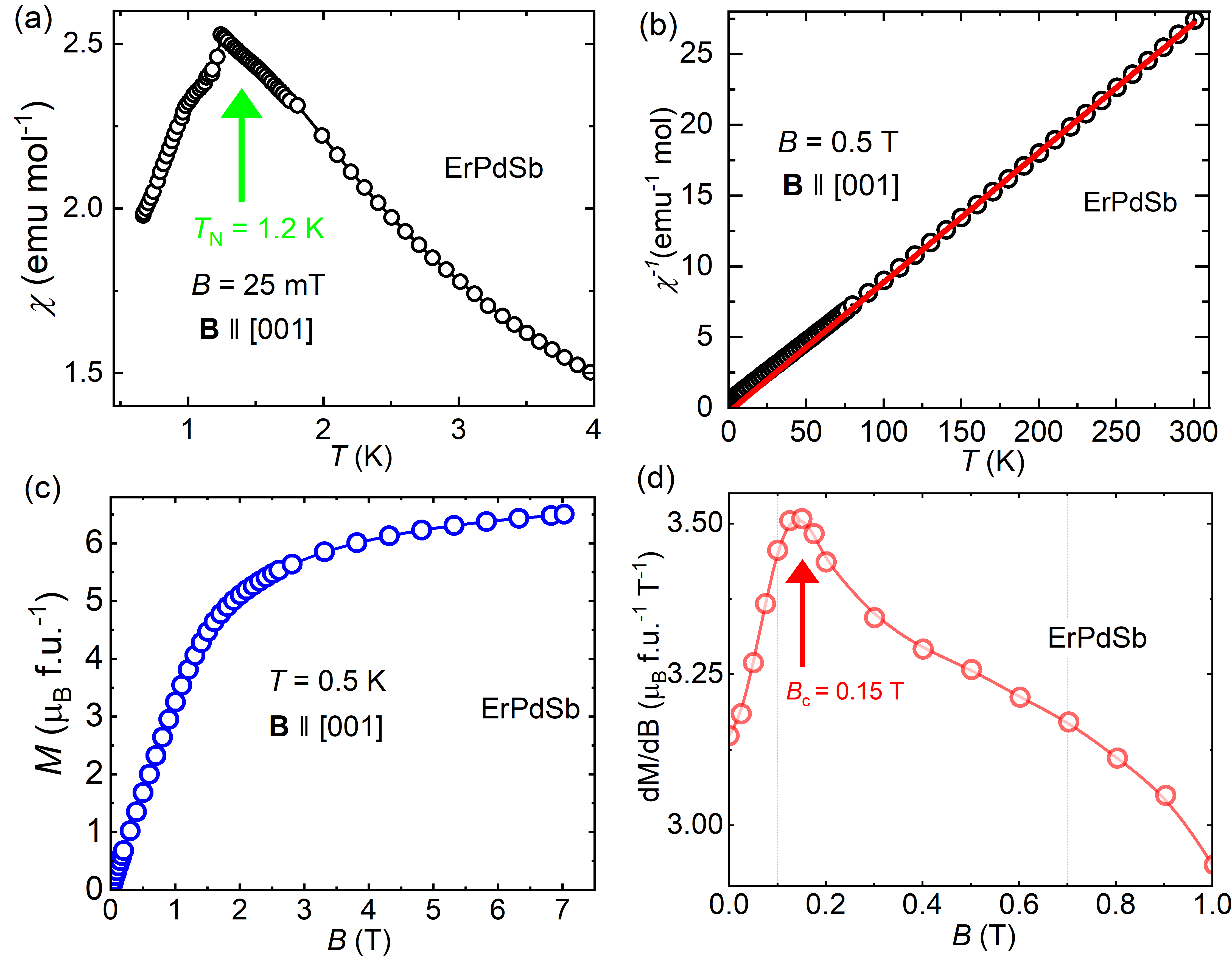}
    \caption{Magnetic properties of single-crystalline ErPdSb. (a) The low-temperature magnetic susceptibility probed in a small field of 25~mT directed along the [001] axis. The arrow marks the antiferromagnetic phase transition. (b) Temperature dependence of the reciprocal magnetic susceptibility measured in a magnetic field of 0.5 T applied along the [001] direction. The red straight line represents the Curie-Weiss fit discussed in the text.  (b) Magnetic field variation of the magnetization measured at 0.5 K in magnetic field applied along the [001] direction. (d) The field derivative of the magnetization versus field. The arrow marks a metamagnetic-like transition.}
    \label{Mag}
\end{figure}

Fig. \ref{Mag}(a) presents the temperature dependence of the magnetic susceptibility $\chi (T)$ of single-crystalline ErPdSb measured in a magnetic field directed along the crystallographic [001] axis. In a small field of 25~mT, $\chi (T)$ exhibits a distinct peak at $T_{\rm N}$ = 1.2 K characteristic of the AFM phase transition. The finding of AFM in ErPdSb corroborates the predictions based on the thermodynamic data collected above 1.7 K \cite{Gofryk2007}. Down to 100 K, the inverse magnetic susceptibility $\chi^{-1}(T)$ follows the Curie-Weiss (CW) law with the effective magnetic moment $\mu_{\rm {eff}}$ = 9.35 $\mu_{\rm B}$ and the paramagnetic Curie-Weiss temperature $\theta_{\rm p}$ = 3.2 K. At lower temperatures, $\chi^{-1}(T)$ slightly deviates from a straight-line behavior (Fig. \ref{Mag}(b)), signaling the effect of crystalline electric field (CEF) interaction. The experimental value of $\mu_{\rm {eff}}$ is fairly close to the theoretical prediction for a free Er$^{3+}$ ion (9.59 $\mu_{\rm B}$). In turn, the small positive value of $\theta_{\rm p}$ hints at the predominance of ferromagnetic-like exchange interactions. It is worth noting that similar CW parameters were reported for polycrystalline samples of ErPdSb ($\mu_{\rm {eff}}$ = 9.43 $\mu_{\rm B}$ and $\theta_{\rm p}$ = -4.2 K \cite{Gofryk2005}). The AFM nature of the electronic ground state in this material is further confirmed by the behavior of the field-dependent magnetization measured at $T$ = 0.5 K (see Fig. \ref{Mag}(c)). The isotherm has a Brillouin-type shape, and it is fully reversible, i.e., shows no hysteresis or remanence effects. At 0.15~T, $M(B)$ undergoes a faint inflection (Fig. \ref{Mag}(d)), which can be attributed to a metamagnetic-like transition. In stronger fields, the magnetization is a curvilinear function of $B$ due to field-governed gradual alignment of the magnetic moments in the field direction. In $B$ = 7 T, the magnetization equals 6.44 $\mu_{\rm B}$, which is much smaller than the saturated magnetic moment of free Er$^{3+}$ ions (9 $\mu_{\rm B}$). Similar reduction was observed before for polycrystals of ErPdSb \cite{Gofryk2007}, and arises due to the CEF effect. The magnetization data measured at higher temperatures are shown in SM\cite{SuppMat} (see Fig. S2). With increasing temperature, the $M(B)$ variations gradually tend towards a straight-line behavior, as expected for paramagnetic materials. 

\begin{figure}[b]
    \centering
    \includegraphics[width=1\linewidth]{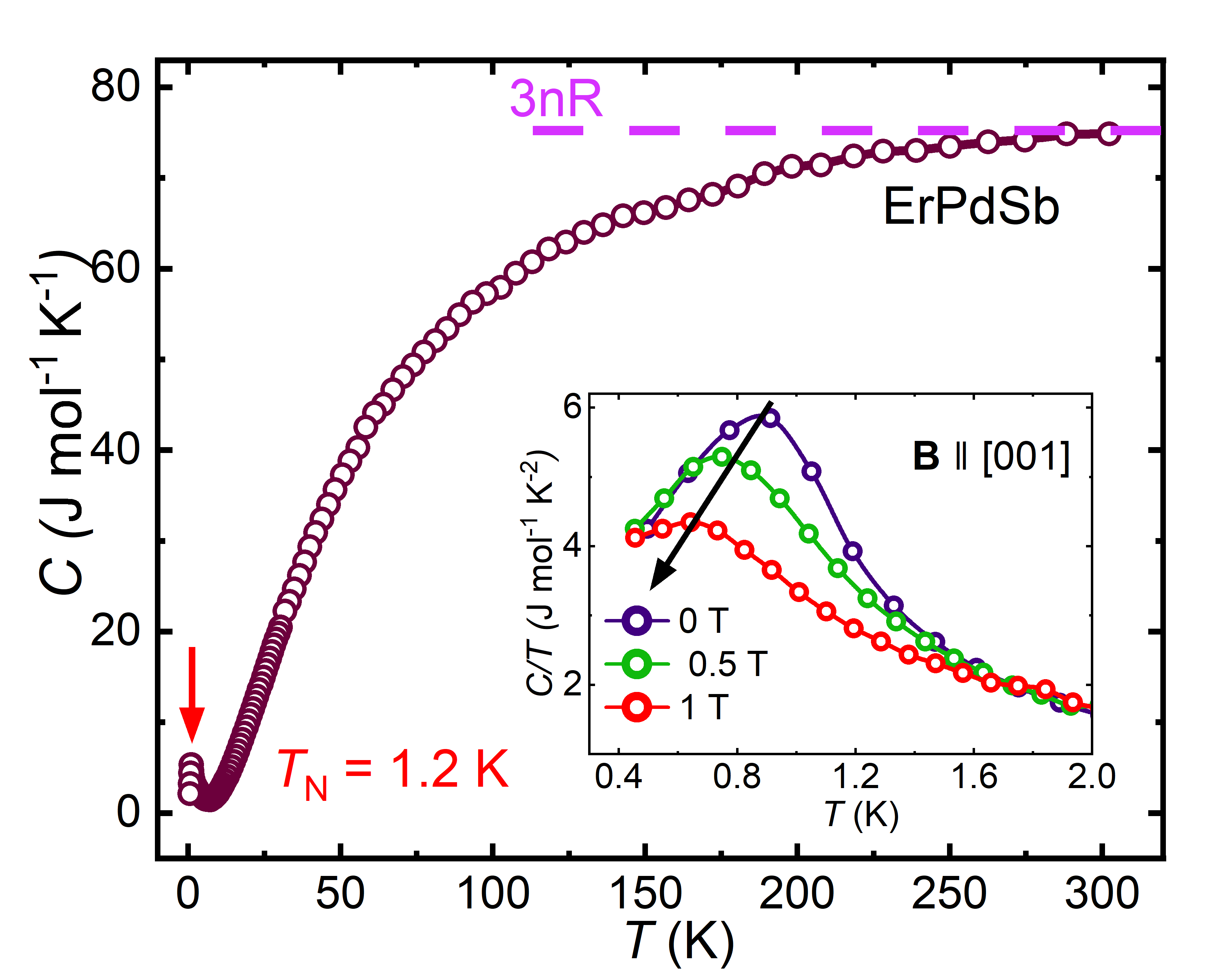}
    \caption{Temperature dependence of the specific heat of single-crystalline ErPdSb. The horizontal dashed line marks the Dulong-Petit limit. The arrow marks the antiferromagnetic phase transition. Inset: the ratio of specific heat over temperature measured at low temperatures in zero and finite magnetic fields applied along the [001] axis.}
    \label{Heat}
\end{figure}

Fig. \ref{Heat} shows the temperature dependence of the specific heat of single-crystalline ErPdSb. The $C(T)$ data obtained above 2 K is in very good agreement with that reported before \cite{Gofryk2007}. Near room temperature, $C(T)$ saturates at the Dulong-Petit limit 3$nR$, where $n$ is the number of atoms in the formula unit, and $R$ stands for the gas constant. With decreasing temperature, $C(T)$ has a  sigmoid-like shape reflecting the predominance of the lattice contribution. However, below about 4 K, the specific heat starts to rise, and forms a lambda-like peak at the onset of the AFM transition (see the inset to Fig. \ref{Heat}). An equal entropy construction yielded an estimate $T_{\rm N}$ = 1.2 K, in concert with the Neel temperature determined from the $\chi(T)$ data. The AFM nature of this phase transition was corroborated by measuring the specific heat in external magnetic field. As can be seen in the figure, as the magnetic field increases, the peak in $C/T$ shifts towards lower temperatures, which is typical for antiferromagnets. 

\subsection{Magnetotransport properties}
\begin{figure*}[t]
         \includegraphics[width=\linewidth]{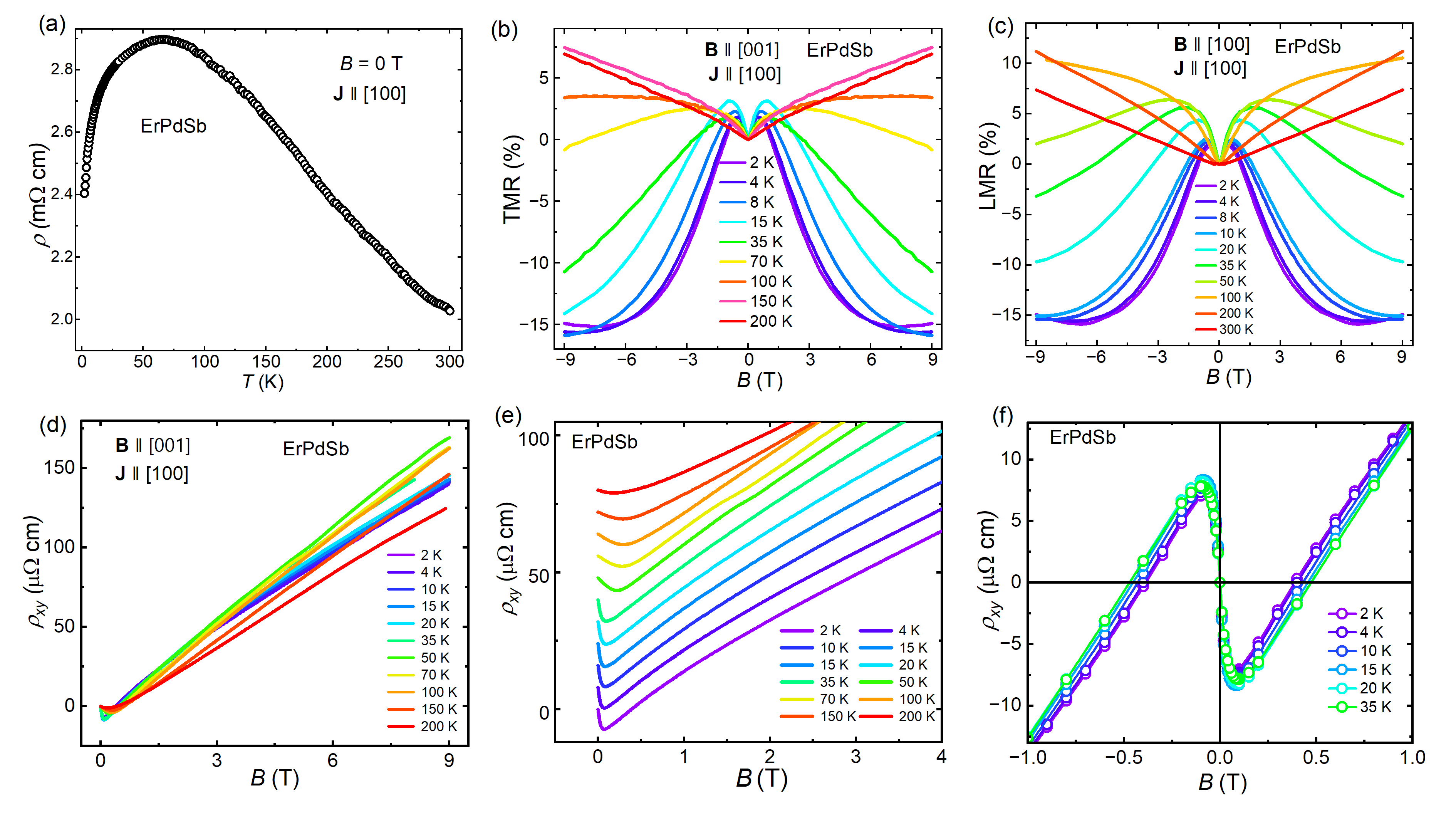}
         \caption{Electrical transport properties of single-crystalline ErPdSb. (a) Temperature dependence of the electrical resistivity measured with the electric current flowing along the crystallographic [100] direction. (b) Transverse magnetoresistance isotherms. (c) Longitudinal magnetoresistance isotherms measured in different magnetic fields applied parallel to the electric current flowing along the [100] axis. (d) Magnetic field dependence of the Hall resistivity measured at several temperatures, with the current flowing along the [100] direction and the magnetic field applied along [001] direction. 
         (e) The Hall resistivity at different temperatures (curves for $T>2$\,K are vertically offset by multiples of $8\,\mu\Omega$\,cm).
         (f) The low-temperature Hall resistivity measured as in panel (d) in weak magnetic fields. }
         \label{res}
\end{figure*}

Fig. \ref{res}(a) displays the temperature dependence of the electrical resistivity, $\rho(T)$, of single-crystalline ErPdSb. Both the magnitude and the shape of $\rho(T)$ indicate a semimetallic or narrow-gap semiconducting character of the material studied. With decreasing temperature, the resistivity first slightly increases, forms a broad  hump in $\rho(T)$ around 70 K, and then begins to decrease. This behavior is fairly similar to that observed in polycrystalline samples studied in Ref. \onlinecite{Gofryk2007}, except for the hump location, which was at much lower temperatures, namely around 20\,K. In contrast, the more recent polycrystalline data reported in Ref.\,\onlinecite{Mukhopadhyay2018} indicated a semiconducting behavior of ErPdSb in the entire temperature range covered. It should be noted that the resistivity of the sample studied in that work was two orders of magnitude larger that that presented in Fig. \ref{res}(a) and reported in Ref.\,\onlinecite{Gofryk2007}. The electrical resistivity is known to be highly sensitive to both the quality of the sample and its exact chemical composition, and this sensitivity is particularly crucial in low-carrier systems. It has previously been shown for several HH phases that the $\rho(T)$ curves demonstrate strong sample dependence, even for the same material. For example, in the case of GdPtBi, one of the most extensively studied half-Heusler topological semimetal, the $\rho(T)$ variations were reported to range from typical semiconducting-like to semimetallic-like and featuring a broad hump, which was attributed to different type of dominant carriers and variations in their concentrations \cite{Hirschberger2016}. 
Our own recent studies on high-energy electron irradiated samples of GdPtBi and LuPdBi also highlight the influence of impurity scattering on the character of $\rho(T)$ behavior \cite{Pavlosiuk2025, Ishihara2021}. Therefore, the fact that the present study was carried out on the high-quality single crystals of ErPdSb, in contrast to the previously studied polycrystalline samples \cite{Mukhopadhyay2018,Gofryk2007}, likely contributes to the observed differences in $\rho(T)$, as polycrystals typically show larger grain boundary scattering and compositional inhomogeneities than single crystals.

Figures \ref{res}(b) and \ref{res}(c) show the magnetoresistance (MR = $\frac{\rho(B)-\rho(0)}{\rho(0)} \times 100\%$) measured as a function of the magnetic field applied perpendicularly (transverse magnetoresistance; TMR) and parallel (longitudinal magnetoresistance; LMR) to the direction of electric current, respectively. In both configurations, at low temperatures ($T$ < 10~K), MR is initially positive and shows a small maximum near 0.5 T. The behavior observed in low fields may be attributed to the weak antilocalization (WAL) effect, as discussed in Section \ref{discussion}. In stronger fields, TMR and LMR change sign, and their absolute values increase, reaching about -15\% above 6~T. The negative magnetoresistance (nMR) may have different origins, which will be addressed below. At low temperatures, the behavior of TMR measured on single-crystalline ErPdSb is quantitatively different from that reported for polycrystals. Nevertheless, some common features can be noted. In both Ref. \onlinecite{Gofryk2007} and Ref. \onlinecite{Mukhopadhyay2018}, TMR increases sharply in weak magnetic fields, reaching a maximum, after which its magnitude decreases, similar to the behavior observed in the present study. In the case of the sample studied in Ref. \onlinecite{Gofryk2007}, the magnetoresistance tends to saturate in high magnetic fields, whereas for the sample studied in Ref. \onlinecite{Mukhopadhyay2018}, TMR shows a significant positive upturn in high fields. Additionally, at $T$ = 8 and 10~K, the polycrystal studied in Ref. \onlinecite{Gofryk2007} exhibits a small nMR in the strongest magnetic fields applied. 
A common feature among all discussed samples is the complex behavior of TMR, characterized by the coexistence of both positive (increasing with field) and negative (decreasing with field) contributions. The relative dominance of these contributions varies from sample to sample.

At higher temperatures, MR becomes somewhat anisotropic, however, it maintains fairly similar behavior that can be seen as an interplay of positive and negative components, which exhibit opposite changes in their magnitudes with increasing temperature. Above 100 K, i.e. in the region of the negative slope in $\rho(T)$, TMR and LMR are positive and nearly proportional to the applied magnetic field. 

Fig. \ref{res}(d) presents the magnetic field dependence of Hall resistivity, $\rho_{xy}$, measured over the temperature range from 2 to 200~K. The positive slope in $\rho_{xy}(B)$ indicates that holes are the majority charge carriers. 
Using a single band model, the carrier density $\mathrm{n}_h$ was estimated to be of the order of 10$^{19}\,\rm{cm}^{-3}$, consistent with the values reported for other semimetallic HH compounds \cite{Dan2024,Pavlosiuk2019}, and particularly with those found previously for polycrystalline samples of ErPdSb \cite{Gofryk2005, Gofryk2007, Mukhopadhyay2018}. As shown in Fig. \ref{res}(f), in weak magnetic fields, $\rho_{xy}(B)$ shows a faint minimum, the position of which is temperature-independent. A similar feature was previously observed in polycrystalline ErPdSb, although the minimum there appeared in stronger magnetic fields of about 1\,T \cite{Gofryk2007}. 
In Fig. \ref{res}(e), $\rho_{xy}$ is plotted separately with an offset of 8 $\mu\Omega$ cm for magnetic fields up to 4\,T. 
The low-field anomaly persists up to at least 200\,K and is nearly unchanged up to 35\,K, while at higher temperatures slightly shifts to higher fields. This feature may signal  magnetic field-induced changes in the Fermi surface topology, as revealed recently for the HH compounds REAuSn (RE = Ho, Er, Tm), which exhibit Lifshitz transitions in weak magnetic fields \cite{Ueda2025}. 

\begin{figure}[h!]
    \centering
    \includegraphics[width=\linewidth]{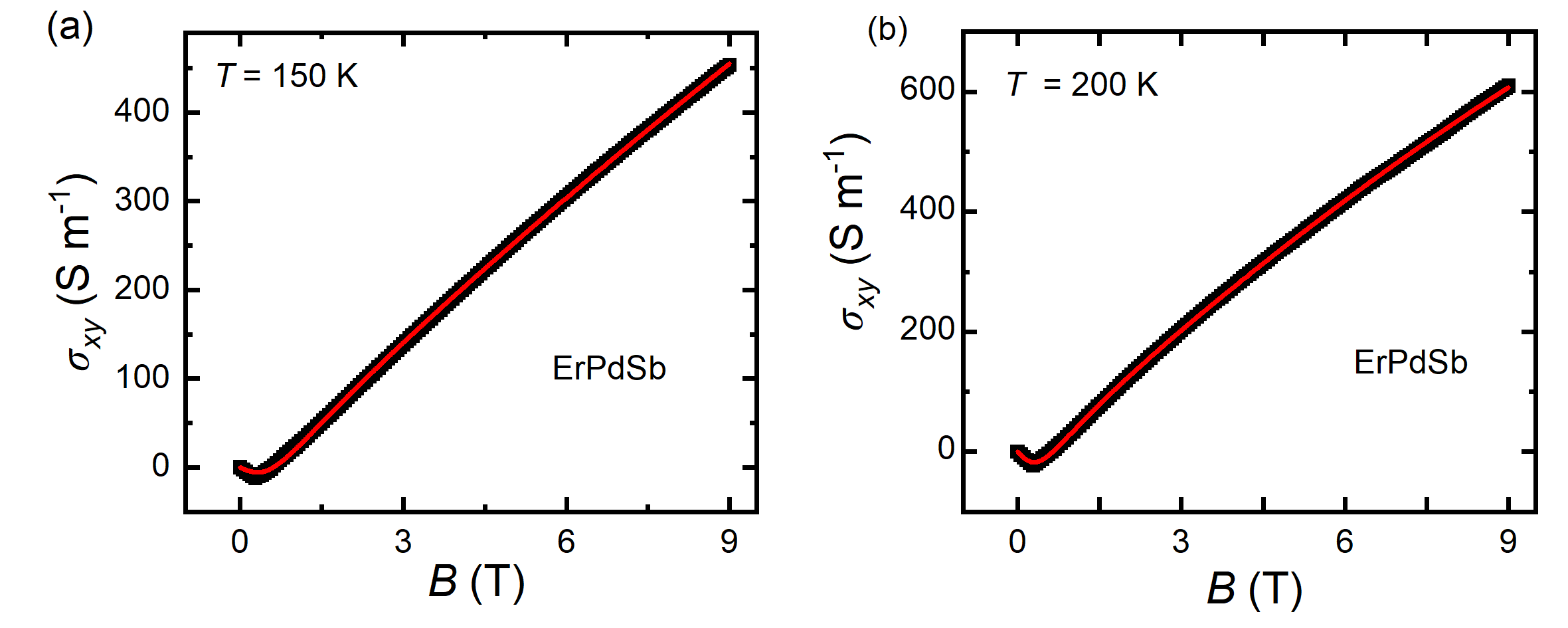}
    \caption{Magnetic field variations of the Hall conductivity in ErPdSb measured as described in Fig. \ref{res}(d) at   (a) 150\,K and (b) 200\,K. The red solid lines show the fits obtained using the two-band model (see the text). }   
    \label{fig:two_b}
\end{figure}

However, the most plausible explanation of the low-field anomaly in the Hall resistivity of ErPdSb is a multi-band nature of this compound. In order to verify this scenario, the Hall conductivity in ErPdSb, derived as $\sigma_{xy}$ = $\frac{\rho_{xy}(B)}{\rho_{xy}(B)^2+\rho(B)^2}$, was fitted in the region $T >$ 100~K using the two-bands model:

\begin{equation}
    \sigma_{xy} = eB\left[\frac{n_h\mu_h^2}{1+(\mu_hB)^2}- \frac{n_e\mu_e^2}{1+(\mu_eB)^2}\right]
    \label{twobandmodel}
\end{equation} 

\noindent where $n_h$ and $n_e$ denote the concentrations of holes and electrons, respectively, while $\mu_h$ and $\mu_e$ stand for the mobilities of these carriers, respectively.
Applying Eq.\,\ref{twobandmodel} to the $\sigma_{xy}(T)$ data measured at $T$ = 150 K and 200 K yielded the curves shown in Fig.\,\ref{fig:two_b}, and the so-derived fitting parameters are listed in Table\,\ref{tab:fitting_parameters}. Clearly, the magnetotransport in ErPdSb has a multiband character with holes being the majority charge carriers. It is very likely, that multiple bands are also involved in the charge conduction at
lower temperatures, however description of the experimental data of ErPdSb collected below $T$ = 100 K in terms of Eq.\,\ref{twobandmodel} was found unreliable due to interdependence of the fitting parameters. 

\begin{table}[h]    
    \caption{
    Hole and electron concentrations ($n_h$ and $n_e$, respectively) and mobilities ($\mu_h$ and $\mu_e$, respectively) in single-crystalline ErPdSb derived from the two-band model applied to the Hall conductivity data measured at $T=150$ and 200\,K.}
    
    \begin{tabular}{ccccc}
        \hline
        T  & $n_h$   & $\mu_h$  & $n_e$   & $\mu_e$ \\
        (K) & ($\rm{m^{-3}}$) & ($\rm{m^{2}V^{-1}s^{-1}}$) & ($\rm{m^{-3}}$) & ($\rm{m^{2}V^{-1}s^{-1}}$)\\
        \hline 
        150   & 3.6 $\times 10^{20}$  & 1.17 & 1.2 $\times 10^{23}$ &  0.020 \\ 
        200  & 3.4 $\times 10^{20}$   &  1.75 & 4.4 $\times 10^{23}$ & 0.032 \\ 
        \hline
    \end{tabular}
    \label{tab:fitting_parameters}
\end{table}
 
\subsection{Angle-dependent magnetoresistance}
\begin{figure*}[hbt]
         \centering
         \includegraphics[width=\linewidth]{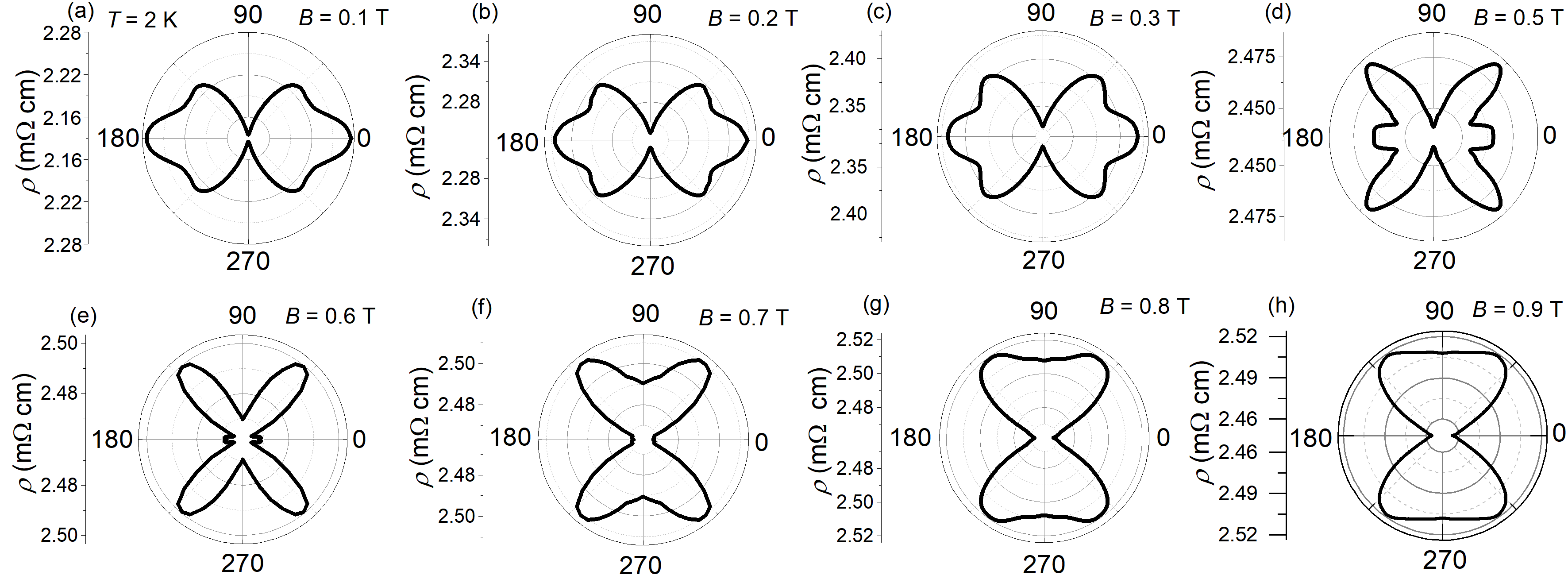}
         \caption{Polar plots of the electrical resistivity of single-crystalline ErPdSb measured at 2\,K in different magnetic fields with the electric current flowing along the [010] direction and the magnetic field rotated within the (010) crystallographic plane.}
         \label{rotxz}
    \end{figure*}

Fig. \ref{rotxz} shows angular variation of $\rho$ at 2~K, in different magnetic fields, there the field is rotated by an angle $\theta$ in the (010) plane and the current flows along the [010] direction. In weak magnetic fields ($B<1$\,T), $\rho$ changes with $\theta$ in a complex manner, but it gradually evolves toward a conventional two-fold symmetry as the field increases (data for $B>1$\,T are shown in SM Fig.\,S5). 
As the magnetic field increases above 0.1\,T, the maximum at 0$^{\circ}$ weakens, while another one present at $\approx45^{\circ}$ becomes stronger, and at 0.5\,T dominates the behavior of $\rho(\theta)$. In stronger fields, a clear maximum in $\rho(\theta)$ appears at $90^{\circ}$ and remains visible up to at least $B = 6$ T (see SM Fig.\,S5). 

To better visualize this complex evolution, we constructed a color map of anisotropic magnetoresistance (AMR), defined as: AMR$(B,\theta)={\frac{\rho(B,\theta) - \rho(B,90^{\circ})}{\rho(B,90^{\circ})}}$ (see the $T=2$\,K data in SM Fig.\,S5(a) and S5(b)). Interestingly, a characteristic magnetic field of 0.6\,T, highlighted by a dashed line, marks a transition point where the peaks and valleys in AMR are inverted. A similar feature was observed before in HoAuSn at 0.8\,T \cite{Lu2025}, and in HoPtBi a similar phenomenon was reported, resulting in the change of AMR symmetry \cite{Chen2023}.
 
Following the procedure applied before to other HH systems \cite{Pavlosiuk2019, Chen2023, Lu2025}, the $\rho(\theta)$ data measured at $T$ = 2\,K in different magnetic fields were analyzed using the formula:

\begin{equation} 
\label{symm}
    \rho(\theta) = C_{0} + C_{2}\cos2\theta + C_{4}\cos4\theta + C_{6}\cos6\theta +C_{8}\cos8\theta
\end{equation}

\noindent where the terms $C_2\mathrm\cos2\theta$, $C_4\mathrm\cos4\theta$, $C_6\mathrm\cos6\theta$, and $C_8\mathrm\cos8\theta$ represent two-, four-, six-, and eight-fold symmetry contributions, respectively. 
The results of such analysis of the anisotropic magnetoresistance of ErPdSb are presented in SM Figs. S5(c) and S5(d), and the field variations of the fitting parameters are in SM Fig.\,S5(e). The most notable feature is the sign change of the $\mathrm{C}_2$ term around 0.6\,T, which reflects a peak-valley inversion in AMR. The $\mathrm{C}_6$ and $\mathrm{C}_8$ are significantly smaller than $\mathrm{C}_2$ and $\mathrm{C}_4$, and they rapidly decrease with increasing magnetic field. 
In the case of HoPtBi, the $\propto\,\cos6\theta$ term was attributed to the 6-fold symmetry of a Fermi pocket centered at $\Gamma$ point \cite{Chen2023}. In turn, the change in AMR symmetry, observed for both HoPtBi and HoAuSn, was associated with changes in the band topology \cite{Chen2023, Lu2025}. It is plausible that similar mechanisms may also be present in ErPdSb.

\section{Discussion }
\label{discussion}
In both measurement geometries, ErPdSb shows a complex magnetoresistance behavior composed of two coexisting components, positive and negative. At low temperatures, a sharp positive cusp appears in fields below 0.5\,T. This low-field behavior can be explained by the weak antilocalization (WAL) effect, which is often observed for topologically non-trivial HH alloys \cite{bhardwaj2018}. Usually, this feature is analyzed in terms of the Hikami-Larkin-Nagaoka (HLN) model \cite{Hikami1980} that was developed for two-dimensional (2D) systems. According to this approach, the electrical conductivity due to WAL can be described by the formula:
    \begin{equation}
       \Delta\sigma = \alpha\frac{e^2}{2\pi^2\hbar}\left[\Psi\left(\frac{1}{2}+\frac{\hbar}{4eL{_\phi^2}B}\right)-\ln\left(\frac{\hbar}{4eL_{\phi}^2B}\right)\right]
    \label{HLN}
    \end{equation}
\noindent where the prefactor $\alpha$ is related to the strength of SOC, $L_{\phi}$ is the phase coherence length, and $\Psi$ is the digamma function. 

\begin{figure}[h]
    \centering
    \includegraphics[width=\linewidth]{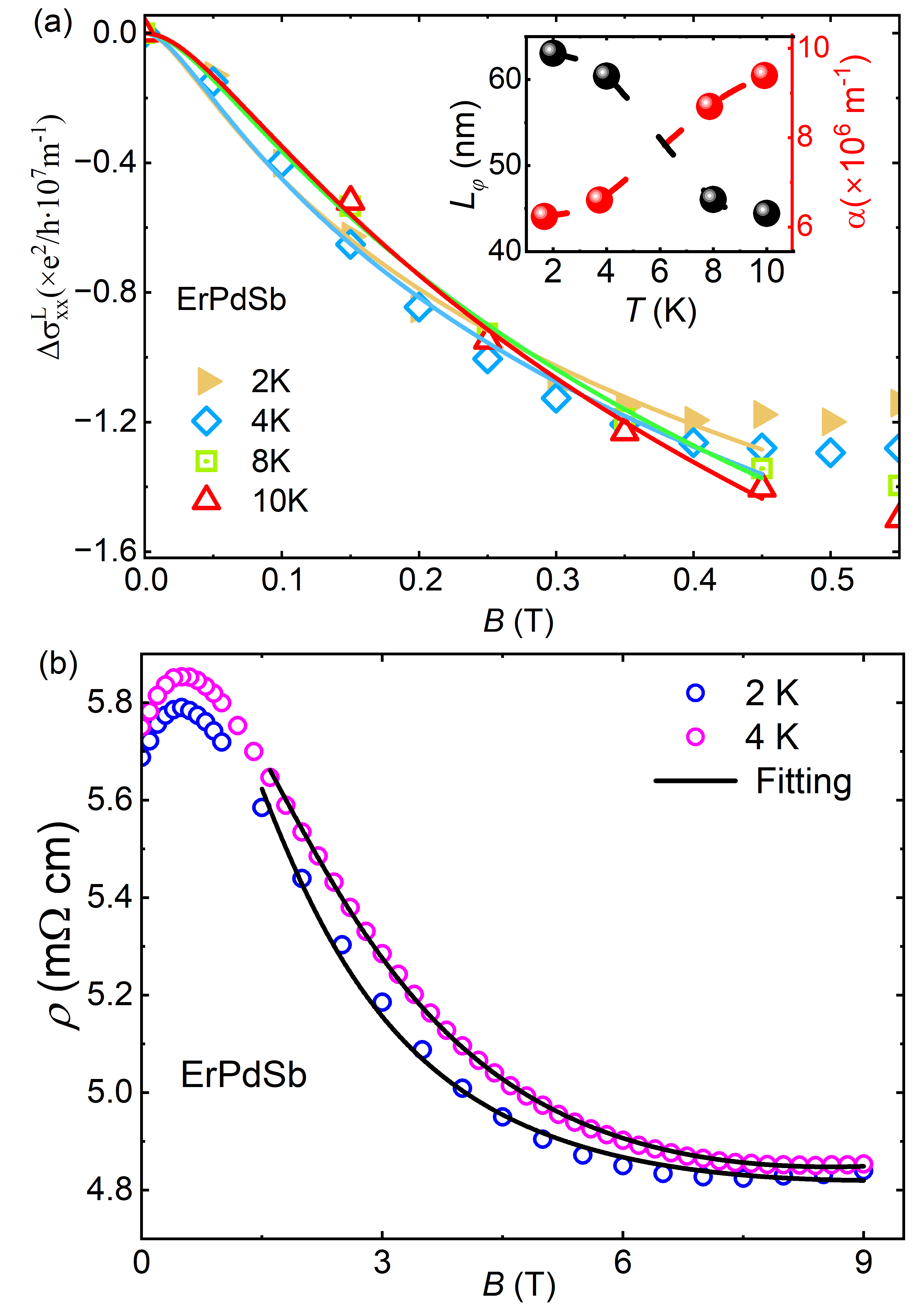}
    \caption{(a) The weak-field longitudinal conductivity of ErPdSb calculated from the data shown in Fig. \ref{res}. The solid lines represent the curves obtained from the HLN fitting (see the text). Inset: temperature variations of the HLN parameters ${\alpha}$ and ${L_{\phi}}$. (b) Magnetic field dependencies of the transverse magnetoresistivity of ErPdSb measured as described in Fig. \ref{res} at $T$ = 2 and 4 K, and their approximations (solid lines) using an approach assuming the dGF and Drude scattering mechanisms (see the text).}
    \label{fig:WAL}
\end{figure}

Fitting Eq.\,\ref{HLN} to the magnetoconductivity data obtained for ErPdSb at $T$ = 2\,K in the longitudinal measurement geometry (see Fig.\,\ref{fig:WAL}(a)) yielded $\alpha$ of the order of $\sim 10^6$, and $L_{\phi}$= 64\,nm. In 2D materials, $\alpha$ typically ranges between 0.4 to 1.1 \cite{Checkelsky2011}, suggesting that one or two surface bands contribute to the electrical conduction, regardless of the presence of other multiple carrier channels from the bulk and surface states \cite{Lu_2014}. However, in the case of HH phases $\alpha$ is typically similar to that found for ErPdSb, and this finding can be ascribed to multiple contributions from bulk and sidewall conducting channels \cite{Xu2014, Pavlosiuk2016, Dan2024}. The temperature dependence of $L_\phi$ and $\alpha$ is shown in the inset to Fig.\,\ref{fig:WAL}(a). 
With increasing temperature, $L_\phi$ decreases, indicating weakening of the WAL effect. It should be noted that WAL is isotropic and was observed in both TMR and LMR. The temperature dependencies of $\alpha$ and $L_\phi$, derived from applying Eq.~\ref{HLN} to the transverse magnetoconductivity data of ErPdSb is shown in SM Fig.\,S4.

In strong magnetic fields and at low temperatures, the magnetoresistance of ErPdSb is negative and displays similar behavior in both transverse and longitudinal measurement configurations. The origin of nMR, often observed for HH materials, can be attributed to various mechanisms, among which CMA is particularly notable \cite{Lu2017}. An archetypal example is semimetalic GdPtBi, where CMA reflects the appearance of magnetic-field induced Weyl nodes, which results in suppression of LMR \cite{Hirschberger2016}. Most recently, similar behavior of LMR was reported, e.g. for HoAuSn \cite{Lu2025}. In the case of ErPdSb, however, CMA should be ruled out as a dominant origin of nMR, because TMR is also negative in strong magnetic fields. Instead, one can consider a mechanism proposed for the half-topological semimetal TbPdBi, namely the cancellation of Berry curvature between trivial spin-up and non-trivial spin-down bands, which brings about a large isotropic nMR ($\rm{98 \%}$) \cite{Zhu2023}. It is worth recalling that in systems with macroscopic disorder, such as polycrystalline $\mathrm{Ag_{2\pm \delta}Se}$ \cite{Parish2003} or distorted topological insulator $\mathrm{TlBi_{0.15}Sb_{0.85}Te_{2}}$ \cite{Breunig2017}, nMR was ascribed to the distortion of current paths. However, this mechanism can be safely excluded in the present case, as the measured samples were high-quality single crystals, free from such major defects, as confirmed by the EDX analysis and Laue diffraction. 

More plausible explanation of nMR found for ErPdSb is the reduction of spin-disorder scattering due to the alignment of magnetic moments in external magnetic field. This mechanism, described by de Gennes and Friedel (dGF) \cite{Gennes}, was used before to explain nMR in other HH compounds \cite{Karla1998, Pavlosiuk2016}. According to the latter approach, the electrical resistivity varies with magnetic field as $\rho_{\rm {dGF}}(B) = \rho_0 [1-(M(B)/M_s)^2]$, where $\rho_0$ represents the strength of the spin-disorder scattering, $M(B)$ is the field-dependent magnetization that can be approximated by the Brillouin function, and $M_s$ is the saturation magnetization. In addition to the dGF mechanism, one can expect that MR is governed also by classical carrier scattering, usually represented by the Drude term $\rho_{\rm D}(B) = \beta B^2$.

In the present case of ErPdSb, in the first step, the experimentally measured $M(B)$ data were fitted using the Brillouin function, with the experimentally determined value $M_s$ = 6.44 $\mu_{B}$. Then, the experimental $\rho(B)$ data measured at $T$ = 2 and 4 K were fitted with the function
\begin{equation}
    \rho(B) = \rho_{dGF}+\rho_{D} = \rho_0[1-(M(B)/M_s)^2] + \beta B^2
\label{fdg}
\end{equation}
\noindent which was found to provide for them a reasonably good description in a relatively wide magnetic field interval (see Fig.\,\ref{fig:WAL}(b)). The so-derived fitting parameters were: $\rho_0$ = 6.10 m$\ohm$ cm and $\beta$ = 0.00615 at 2~K, and $\rho_0$ = 5.97 m$\ohm$ cm and $\beta$ = 0.0019 at 4 K. Although the contribution from the Drude mechanism is relatively small compared to the dGF spin-disorder scattering, its effect becomes noticeable particularly in strong magnetic fields.

Alternatively, nMR may stem from the field-induced reconstruction of the Fermi surface. According to the bulk electronic structure calculations, ErPdSb is a semiconductor with an indirect gap (see Fig.\,\ref{fig:BAND}a). If one assumes that band splitting due to the Zeeman effect or exchange interactions, similar to the cases of GdPtBi \cite{Hirschberger2016, Suzuki2016} or REAuSn \cite{Ueda2025}, also occurs in ErPdSb,  then the following scenario may take place: at a certain value of the magnetic field, the Fermi level starts to intersect the hole-type band (located around the $\Gamma$ point). In a higher magnetic field, the Fermi level may intersect both the hole-type band and the electron-type band, located around the X point. Consequently, the number of carriers of both types will increase, leading to a decrease in resistivity with increasing magnetic field. Moreover, such a Fermi surface reconstruction may manifest itself in a complex angular dependence of MR, reflecting a strongly anisotropic, ellipsoidal shape of the electron Fermi pocket, which may appear around the X point. Interestingly, the crossover from positive to negative MR occurs in ErPdSb around $B$ = 0.6 T, which coincides with the magnetic field value at which the peaks and valleys in AMR are inverted (cf. Fig.\,S5(b) in SM).

\section{Conclusions}
In summary, we have grown single crystals of ErPdSb using flux method and characterized them through detailed thermodynamic and magnetotransport measurements. Magnetic and heat capacity measurements reveal antiferromagnetic ordering below $T_\mathrm{N}$ = 1.2~K. Isothermal magnetization measurement at 0.5~K shows a faint metamagnetic transition in a weak magnetic field of 0.15~T. The temperature dependence of resistivity in zero magnetic field exhibits semimetallic behavior. The positive magnetoresistance in weak magnetic fields indicates the presence of WAL. At high magnetic fields, the negative magnetoresistance is well described by the de Gennes-Friedel model. Furthermore, the angular dependence of magnetoresistance in the $xz-$plane shows anomalous behavior in weak magnetic fields, with peak-valley inversion at 0.6~T. In the high magnetic fields, the two-fold symmetry dominates in $\rho(\theta)$. Hall resistivity shows non-linear behavior in small magnetic fields, forming a dip below 1~T. This anomaly persists up to 200~K, suggesting a multiband contribution in the transport. We have fitted Hall conductivity at 150~K and 200~K using the two-band model.  
The electronic structure calculations indicated an indirect band gap ($\Gamma$-$X$) of $\approx$0.15 eV and pronounced valence band splittings related to strong spin-orbit coupling in the bulk material. 

\begin{acknowledgments}
This work was supported by National Science Centre (Poland) under project No. 2021/40/Q/ST5/00066.
\end{acknowledgments}

\section*{DATA AVAILABILITY}
Data supporting this study are available upon reasonable request from the corresponding author.

\bibliography{mybib}

\newpage

\onecolumngrid

\begin{center}
  \textbf{\Large Supplementary Material}\\[.2cm]
  \textbf{\large Anomalous magnetotransport in single-crystalline half-Heusler antiferromagnet ErPdSb }\\[.2cm]
  %%%%%%
  Abhinav Agarwal, S. Dan, Maciej J. Winiarski, O. Pavlosiuk, P. Wi\'{s}niewski, and D. Kaczorowski\\[.2cm]
  %%%%%%
  {\itshape
  	\mbox{Institute of Low Temperature and Structure Research, Polish Academy of Sciences, Okólna 2, 50-422 Wroclaw, Poland}\\
	
  }

\end{center}

\setcounter{equation}{0}
\renewcommand{\theequation}{S\arabic{equation}}
\setcounter{figure}{0}
\renewcommand{\thefigure}{S\arabic{figure}}
\setcounter{section}{0}
\renewcommand{\thesection}{S\arabic{section}}
\setcounter{table}{0}
\renewcommand{\thetable}{S\arabic{table}}
\setcounter{page}{1}

\begin{figure}[h!]
    \centering
    \includegraphics[width=0.5\linewidth]{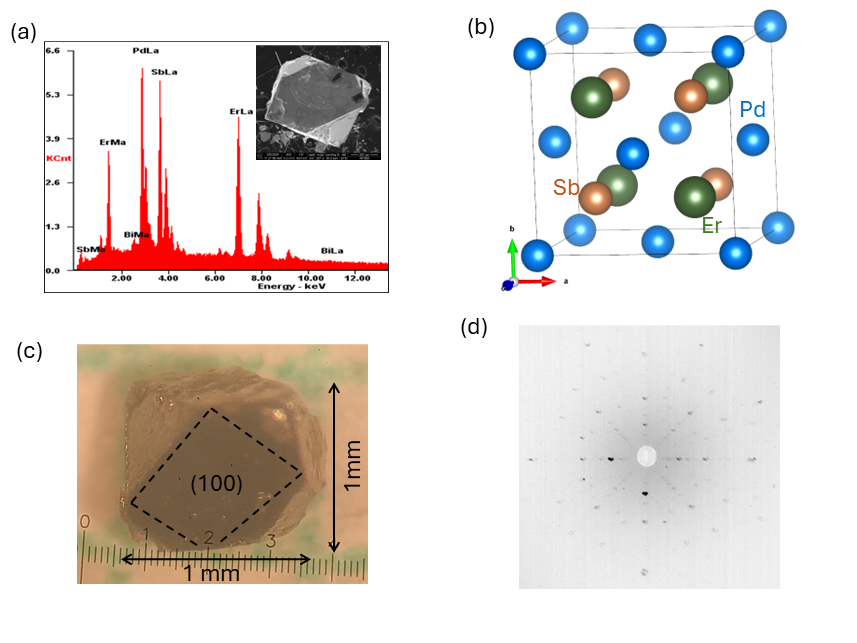}
    \caption{ Fig.(a) Exemplary EDS spectrum of ErPdSb.  Inset: SEM image of the crystal examined.  (b) Model of the crystal structure of ErPdSb drawn using VESTA\cite{momma.izumi.11}.  (c) Optical image of as-grown crystal with upper face being the crystallographic (100) plane. (d) Laue backscattering diffraction pattern taken along the [001] crystallographic direction.}
    \label{fig:edx}
\end{figure}

\begin{figure}[h!]
    \centering
    \includegraphics[width=0.5\linewidth]{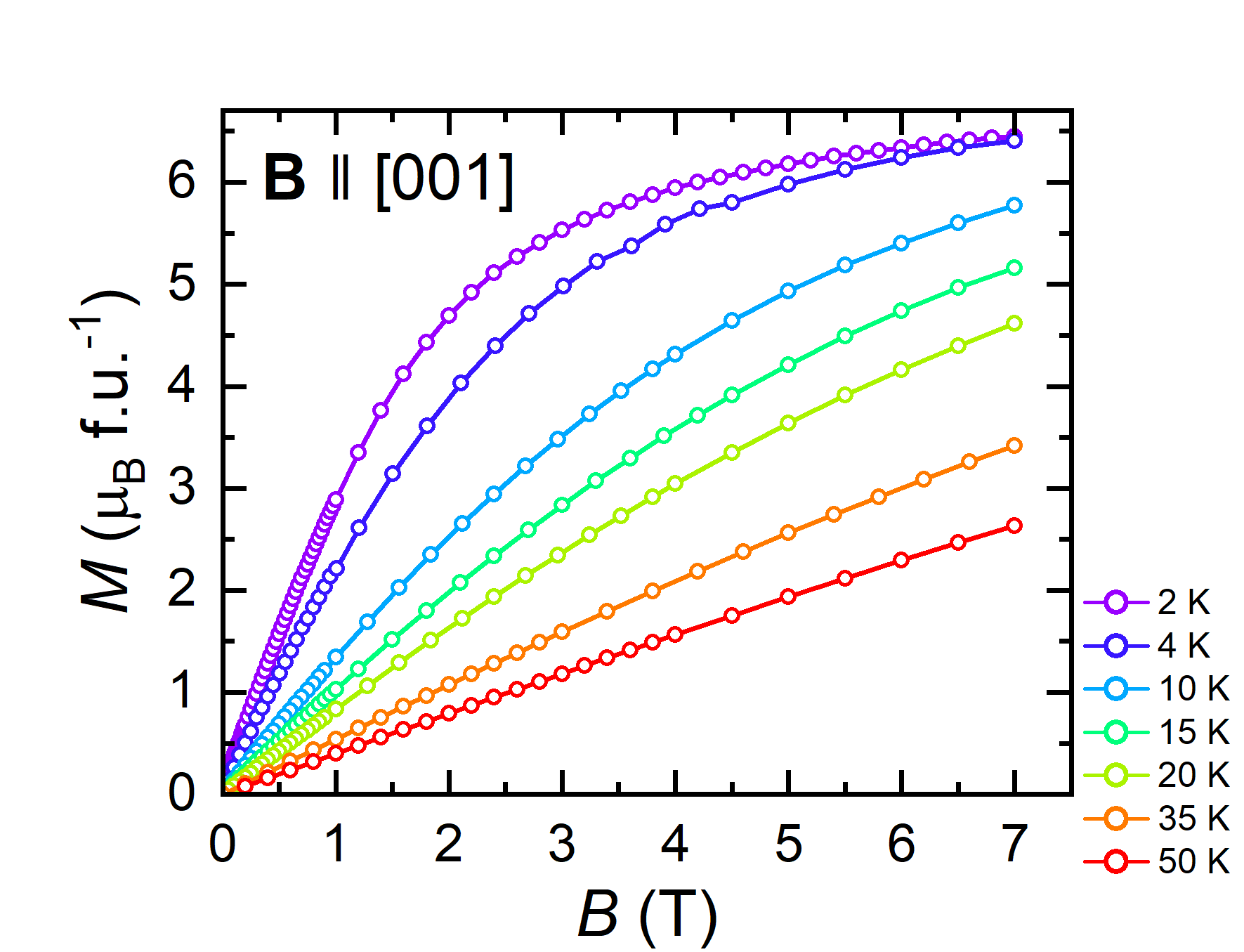}
    \caption{Magnetic field dependencies of the magnetization in ErPdSb measured at $T=$2, 4, 10, 15, 20, 35, and 50 K with magnetic field \textbf{B}$\parallel$ [001].}
    \label{fig:mh}
\end{figure}

\begin{figure}[h!]
    \centering
    \includegraphics[width=0.5\linewidth]{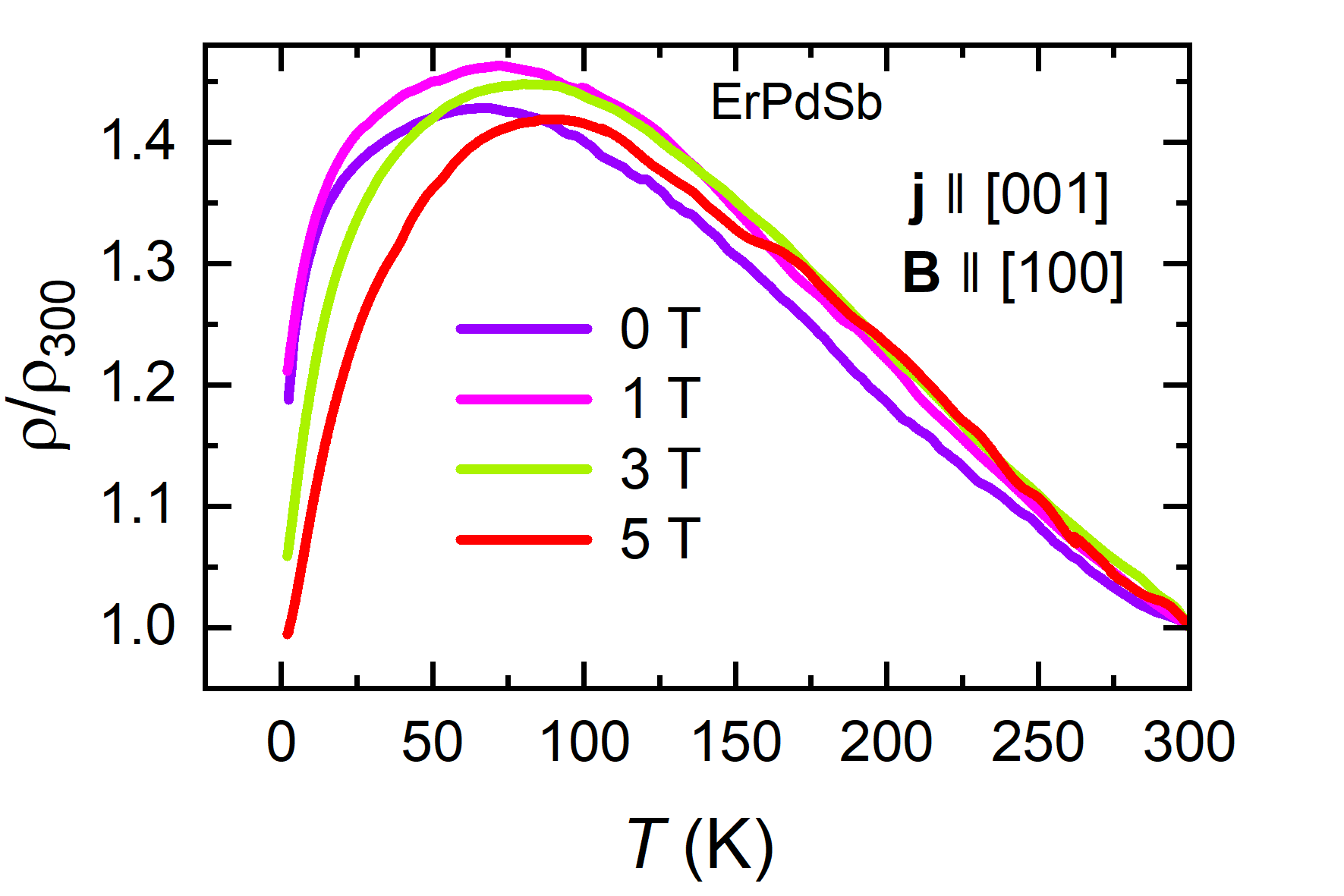}
    \caption{Temperature variations of the electrical resistivity of ErPdSb measured in different magnetic fields and normalized by dividing by its room temperature value.}
    \label{fig:rho}
\end{figure}

\begin{figure}[h!]
    \centering
    \includegraphics[width=0.8\linewidth]{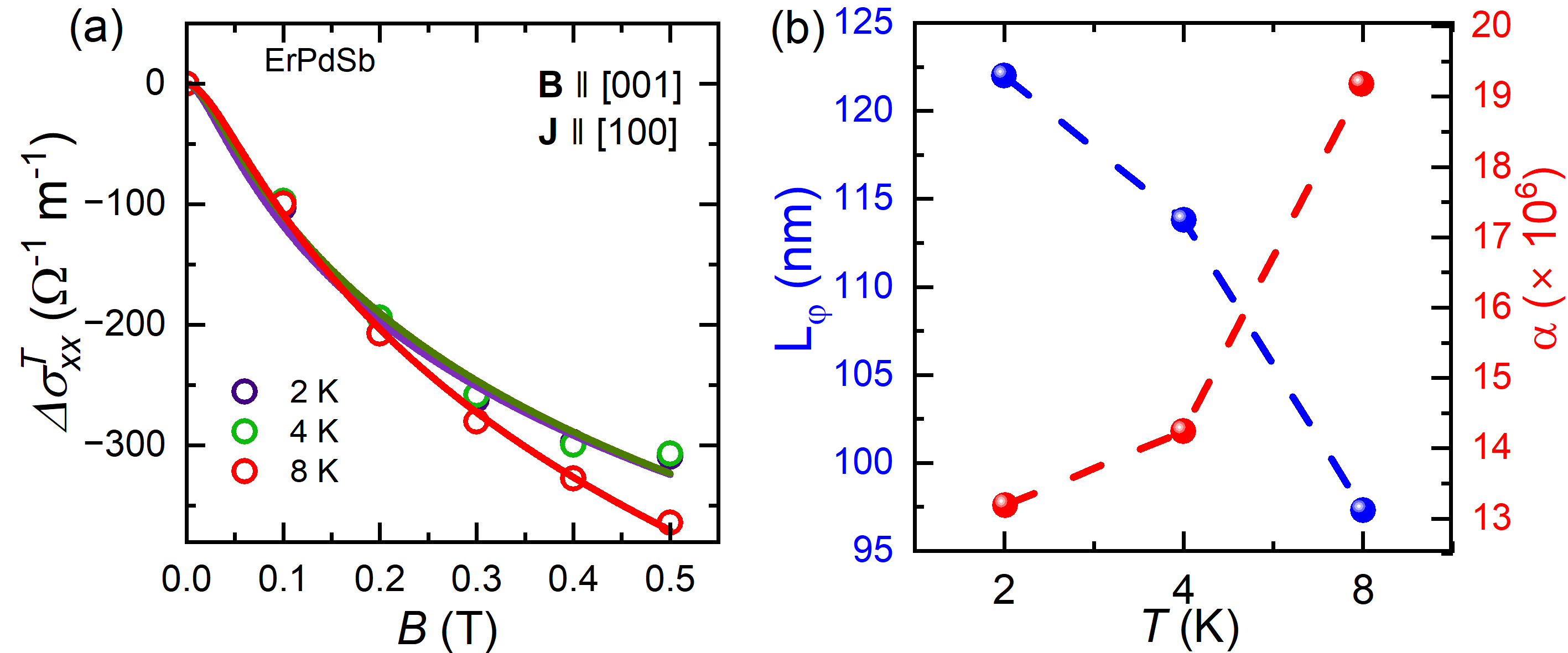}
    \caption{(a) Weak antilocalization fitting of the transverse magnetoconductivity in ErPdSb in terms of the HLN approach. (b) Temperature dependencies of the HLN parameters.}
    \label{fig:wal}
\end{figure}

\begin{figure}[h!]
    \centering
    \includegraphics[width=0.8\linewidth]{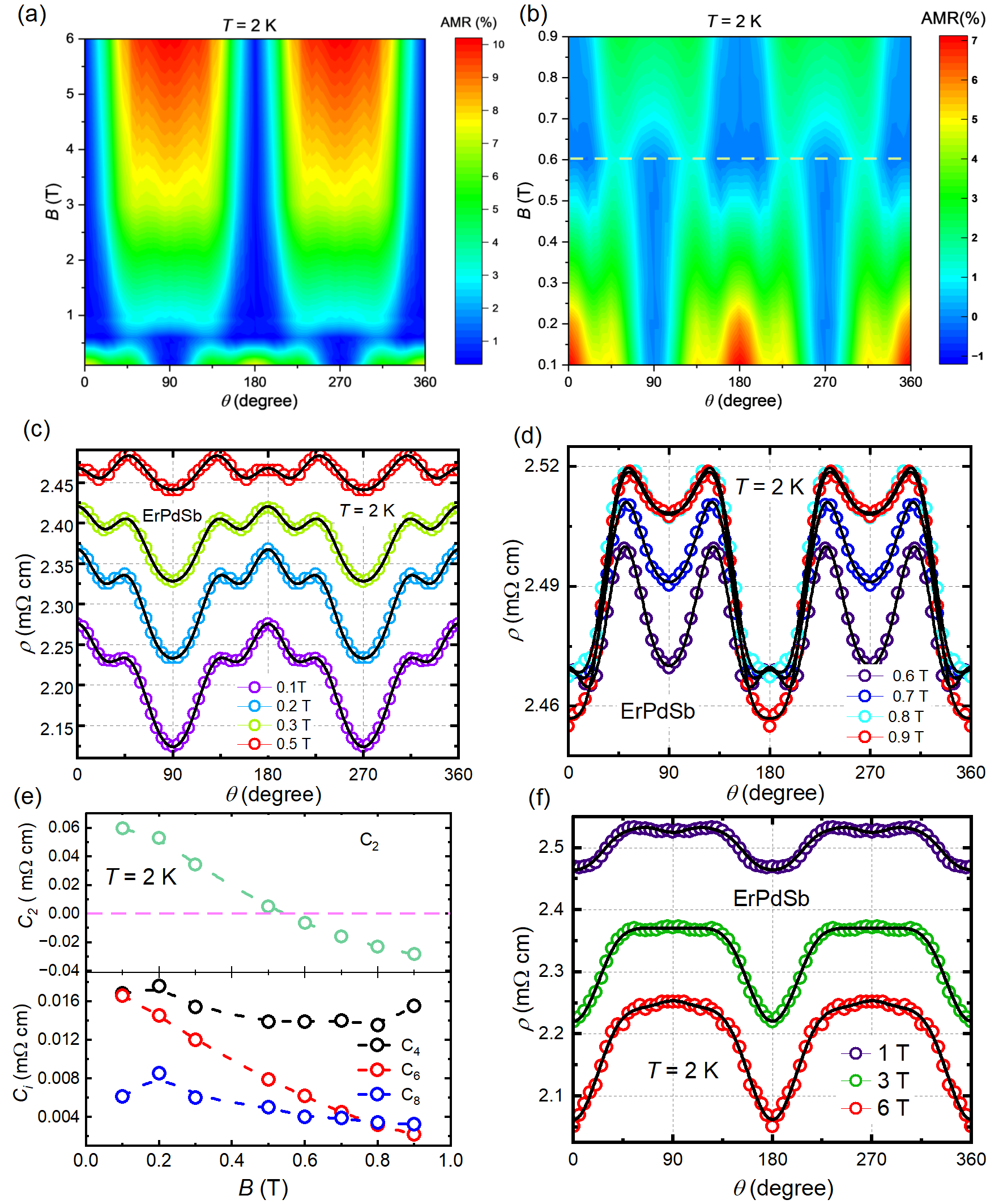}
    \caption{(a) AMR colormap at 2 K for different magnetic fields. (b) AMR colormap at 2 K for small magnetic fields (< 1 T). The dashed horizontal line signifies the transition of AMR, as described in the main text. (c) \&(d) shows the cartesian plots of $\rho$  vs $\theta$  for small magnetic fields ( < 1 T) at 2 K with corresponding fitting. (e) Contribution of different orders of symmetry extracted from fitting. (d) The variation of  $\rho$ with $\theta$ in high magnetic fields. The solid line corresponds to the fitting.}
    \label{fig:amrxzs}
\end{figure}

\begin{figure}[h!]
    \centering
    \includegraphics[width=0.8\linewidth]{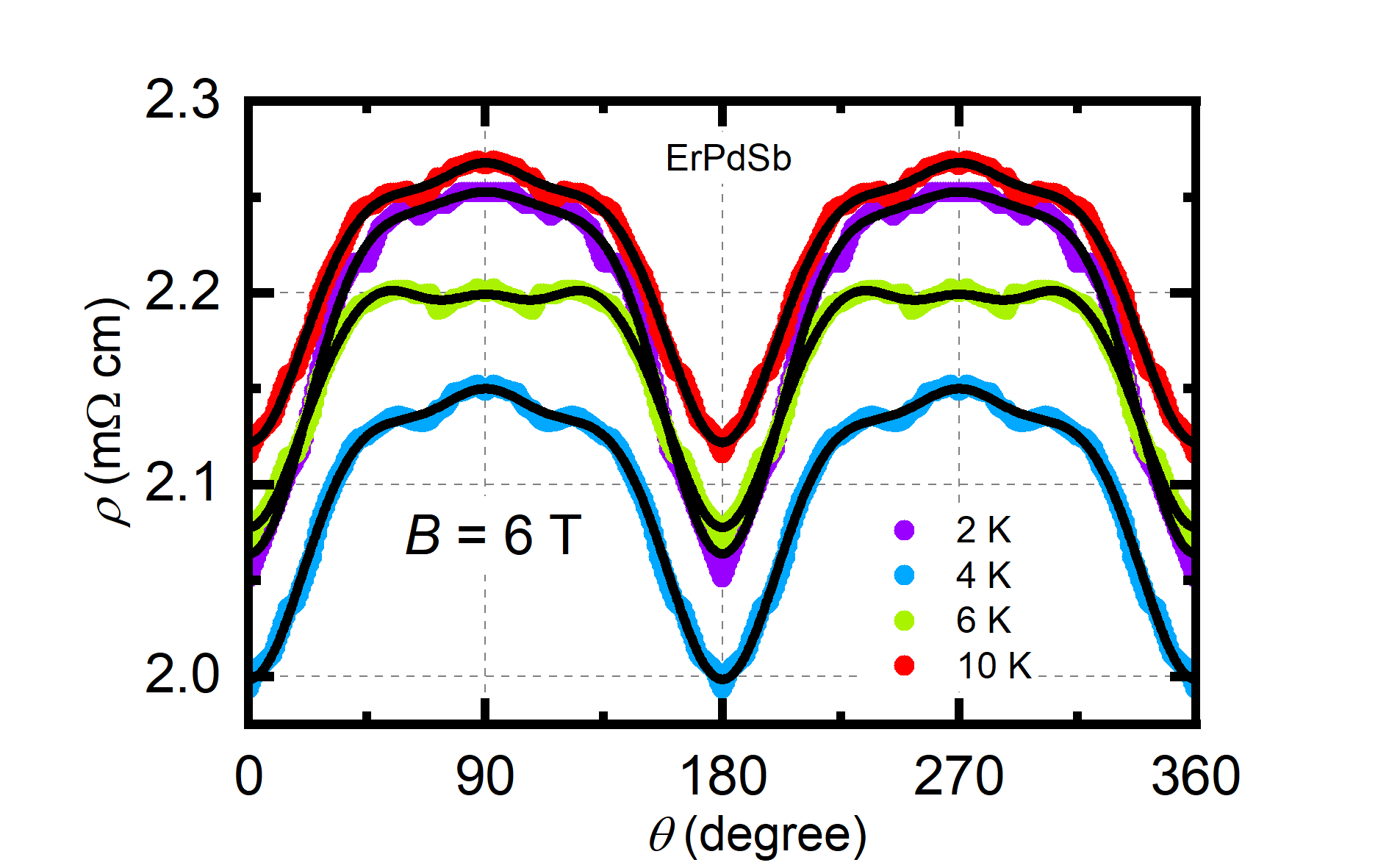}
    \caption{$\rho$ vs $\theta$ at different temperatures in a constant magnetic field of 6~T. The solid line corresponds to the fitting. }
    \label{fig:amrt}
\end{figure}

\begin{figure}[h!]
    \centering
    \includegraphics[width=\linewidth]{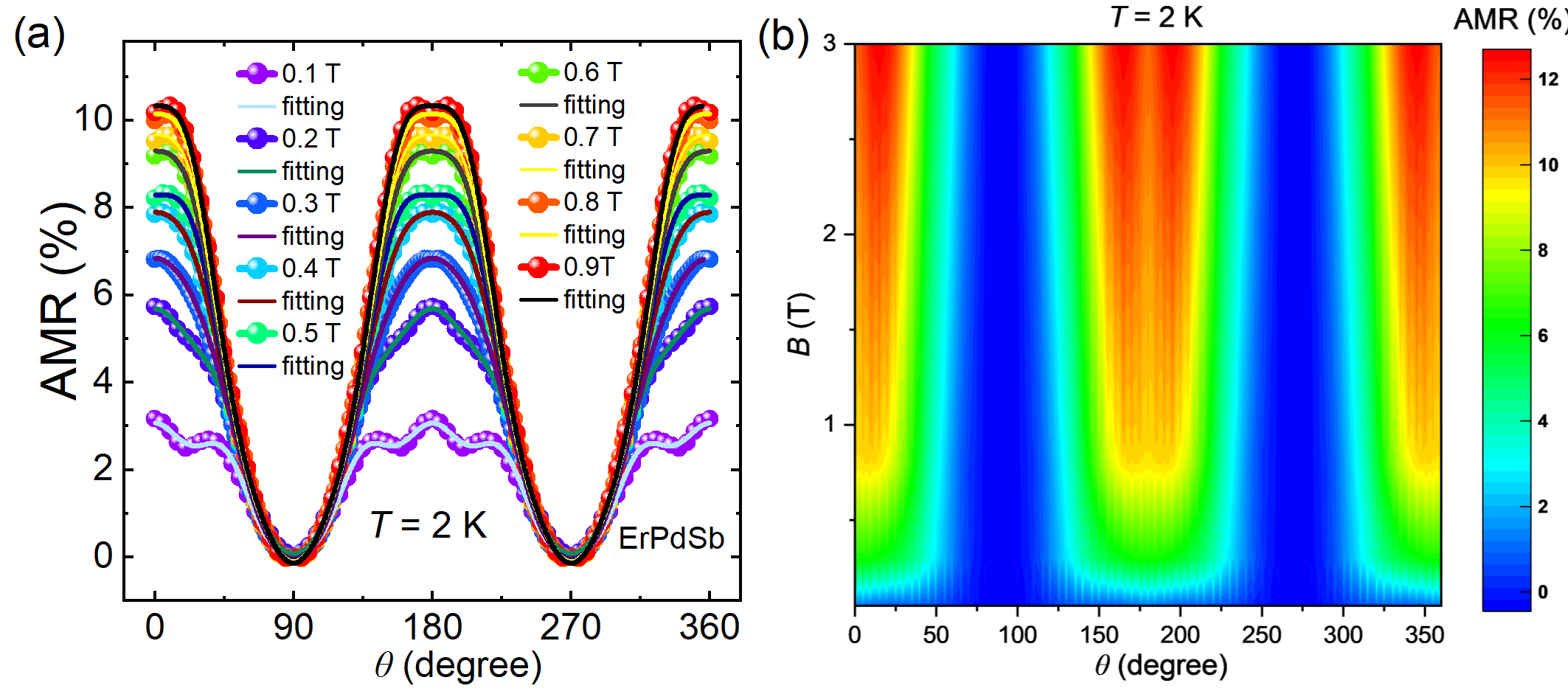}
    \caption{(a) $\rho$ vs $\theta$ at different magnetic fields measured at $T$ = 2~K. The solid line corresponds to the fitting. (b) AMR colormap at 2~K in different magnetic fields.}
    \label{fig:amry}
\end{figure}

In this Supplementary Material, we present the following additional results:
\begin{itemize}
\item Sec.~\ref{sec.crystal} Structural Characterization.

Fig.~\ref{fig:edx} EDS report, optical image \& Laue diffraction pattern

Crystal structure refinement (separate CIF file included)
\item Sec.~\ref{sec.mag} Isothermal magnetization.

Fig.~\ref{fig:mh} Isothermal $M(B)$ curve
\item Sec.~\ref{sec.res} Temperature dependence of resistivity in different magnetic fields.

Fig.~\ref{fig:rho} The $\rho(T)$ variations
\item Sec.~\ref{sec.wal} Weak antilocalization.

Fig.~\ref{fig:wal} WAL fitting for transverse magnetoconductivity
\item Sec.~\ref{sec.amr} Anomalous linear resistivity in different field directions.

Fig.~\ref{fig:amrxzs} AMR for $xz-$plane at 2~K

Fig.~\ref{fig:amrt} AMR for $xz-$plane at 6~T

Fig.~\ref{fig:amry} AMR for $yz-$plane at 2~K
\end{itemize}

\newpage
\section{Structural Characterization}
\label{sec.crystal}

\section{Isothermal $M(B)$ curves}
\label{sec.mag}

\newpage

\section{Temperature dependence of resistivity in different magnetic fields}
\label{sec.res}

\section{WAL fitting for transverse magnetoconductivity}
\label{sec.wal}

\newpage
\section{Anomalous linear resistivity in different field directions}
\label{sec.amr}
In this section, we have shown the angular variation of resistivity ($\rho(\theta)$) in different magnetic fields. The crystal is rotated in the $xz-$plane (\textbf{j} $\parallel$ [100] and \textbf{B} is rotated in the (001) plane) and $yz-$plane (\textbf{j} $\parallel$ [010] and \textbf{B} is rotated in the (110) plane), respectively. For the $xz-$plane, the variation of $\rho(\theta)$ is complex and contains higher-order symmetry terms in the weak magnetic fields. 3D-colormapping of AMR$(B,\theta)={\frac{\rho(B,\theta) - \rho(B,90^{\circ})}{\rho(B,90^{\circ})}}$ with $\theta$ and $B$ is constructed to effectively visualize the change in $\rho(\theta)$ with $B$.
%\section{AMR for xz-plane at 2 K}
\label{sec.amrsz}

%\section{AMR for xz-plane at 6~T}
\label{sec.amrt}

%\section{AMR for yz-plane at 2~K}
\label{sec.amrYZ}

\end{document}